\documentclass[aps,prb,twocolumn,superscriptaddress]{revtex4}
\usepackage{blindtext}
\usepackage{graphicx}

\begin{document}
\title{Percolation of Fortuin-Kasteleyn clusters for the random-bond
Ising model}
\author{Hauke Fajen}
\affiliation{Institut f\"{u}r Physik, Universit\"{a}t Oldenburg, 26111 Oldenburg, Germany}
\author{Alexander K. Hartmann}
\email{a.hartmann@uni-oldenburg.de}
\affiliation{Institut f\"{u}r Physik, Universit\"{a}t Oldenburg, 26111 Oldenburg, Germany}
\author{A. Peter Young}
\affiliation{University of California Santa Cruz, CA 95064, USA}

\begin{abstract}
We apply generalisations of the Swendson-Wang and Wolff cluster
algorithms, which are based on the construction of Fortuin-Kasteleyn clusters,
to the three-dimensional $\pm 1$ random-bond Ising model.
The behaviour of the model is
determined by the temperature $T$ and the concentration $p$ of negative
(anti-ferromagnetic)
bonds.
The ground state
is ferromagnetic for $0 \le p<p_c$, and a spin glass for $p_c < p \le 0.5$
where $p_c \simeq 0.222$.
We investigate the percolation transition of the Fortuin-Kasteleyn
clusters as function of temperature.
Except for $p=0$ the Fortuin-Kasteleyn percolation transition occurs at a
higher temperature than the magnetic ordering temperature.
This was known before for $p=1/2$ but here we provide evidence for a difference
in transition temperatures even for $p$ arbitrarily small.
Furthermore, for all values of $p>0$,
our data suggest that the percolation transition is universal, irrespective
of whether the ground state exhibits ferromagnetic or spin-glass order, and is
in the universality class of standard percolation. This shows that
correlations in the bond occupancy of the Fortuin-Kasteleyn clusters are
irrelevant, except for $p=0$ where the clusters are tied to Ising correlations
so the percolation transition is in the Ising universality class. 

\end{abstract}
\maketitle

\section{Introduction}

Magnetic systems with quenched disorder, such as spin glasses
(SGs) \cite{binder1986,mezard1987,young1998,nishimori2001}
and random field systems,
exhibit
phase transitions between low-temperature ordered and high-temperature
disordered (paramagnetic) phases
in high enough dimensions.
This is similar to the case of pure
systems like ferromagnets \cite{ising1925} but spin glasses
in particular 
exhibit a much richer behaviour and many aspects of the low-temperature
phase are still not well understood.  Since most disordered models
cannot be solved analytically,
one has to resort to computer simulations.\cite{practical_guide2015}
For the special case of zero temperature, there are often
efficient algorithms\cite{opt-phys2001}. However, for 
systems coupled to a heat bath at finite temperature, 
Monte Carlo simulations \cite{landau2000, newman1999} are generally
used.  For the 
\textit{pure} Ising model, efficient cluster Monte Carlo (MC) approaches
exist,\cite{swendsen1987,wolff1989} which are based on the construction
of Fortuin-Kasteleyn (FK) \cite{fortuin1972} clusters of spins.
This gives fast equilibration even
close to the phase transition point. The reason is that 
the FK clusters percolate
\cite{stauffer1994introduction}
\textit{precisely}
at the phase transition.\cite{coniglio1980} 

It is also possible to implement cluster MC algorithms like the Wolff algorithm
for spin glasses, but
unfortunately these are not efficient because, in the vicinity of the spin
glass phase transition, each update
flips almost all the spins.\cite{kessler1990}
The reason is that percolation of
the FT clusters happens at much higher temperatures than the
magnetic-ordering phase transition temperature.\cite{deArcangelis1991}  
Other approaches for cluster algorithms for spin glasses have been tried,
\cite{niedermayer1988,liang1992,houdayer2001,joerg2005,zhu2015} 
but in the end none turned out to be efficient for three-dimensional
spin glasses and related models.
Thus, single-spin flip algorithms are still used for studying spin glasses
numerically. Some improvement is obtained by using parallel
tempering,\cite{geyer1991,hukushima1996} and by running parallel tempering on
a special-purpose high-performance computer ``JANUS'' \cite{janus2009} it
has been possible to simulate an $N=48^3$ spin glass model near the transition
temperature.

To obtain a better understanding of the nature of
FK clusters and their percolation transitions, as well 
as algorithmic efficiency, we
study here the $\pm 1$ random-bond Ising model,\cite{kirkpatrick1977} which is a
generalisation of the standard spin glass. It consist of $N$ Ising
spins $\sigma_i=\pm 1$ placed on a $d$-dimensional hyper-cubic lattice
of linear size $L$, i.e. $N=L^d$. The Hamiltonian is given by
\begin{equation}
H =
- \sum_{\langle i,j\rangle} J_{ij} \sigma_i \sigma_j \,.
\end{equation}
Each spin $i$ interacts with its nearest neighbours $j$ via an
interaction which is a quenched random variable
$J_{ij}$.  Here we use a bimodal distribution so each bond is
anti-ferromagnetic ($J_{ij}=-1$) with probability $p$ and
ferromagnetic ($J_{ij}=+1$) with probability $1-p$. As usual for
quenched disorder, the result of any measurement will depend on
the realisation of the disorder, so one has to perform an average over many
realizations of disorder
in addition to doing the thermal average.

\begin{figure}[htb]
\includegraphics[width=\columnwidth]{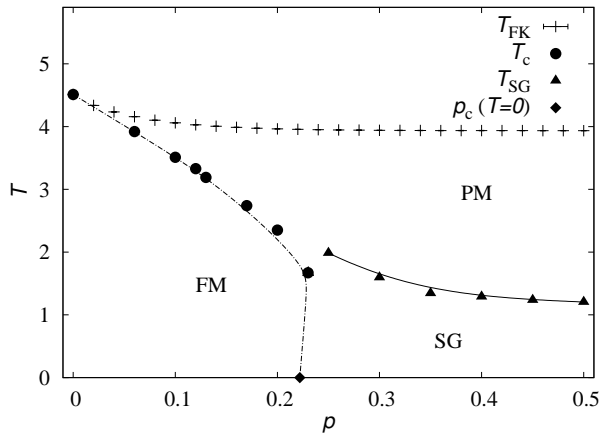}
\caption{\label{fig:phasediagram}
Phase diagram showing the line of percolation transitions of the FK clusters,
and the lines of phase transitions between ferromagnetic (FM), paramagnetic
(PM) and spin glass (SG) phases.  Lines are  guides for the eyes only. The
data for the ferromagnetic transition temperature $T_c(p)$ is from
Ref.~[\onlinecite{hasenbusch2007}], the data for the spin glass transition
temperature $T_{\rm SG}$ is from Ref.~[\onlinecite{reger1986}], and the value
of $p$ where the spin glass and ferromagnetic phases meet at $T=0$, $p_{\rm
c}$, is from Ref.~[\onlinecite{art_threshold1999}]. The data for the
percolation transition temperature $T_{FK}(p)$ is from this work.
}
\end{figure}

We consider here the case of a simple cubic lattice for which the low
temperature phase is ferromagnetic for a small concentration ($p$) of
anti-ferromagnetic bonds and a spin glass for a larger concentration. We
denote the paramagnet to ferromagnet transition temperature by $T_c(p)$ and
the paramagnet to spin glass transition by $T_{\rm SG}(p)$.
The phase diagram in the $p$--$T$ plane has been determined by
Monte Carlo simulations, \cite{reger1986,hasenbusch2007} see
Fig.~\ref{fig:phasediagram}.
For $T=0$ the transition point between the
ferromagnetic and spin-glass phases was found \cite{art_threshold1999}
to be approximately $p_c=0.222(5)$.

In the present study, we investigate the behaviour of FK
clusters and, related to this, the performance of the Wolff algorithm, in the
$p$-$T$ plane.  We know that for the pure ($p=0$) ferromagnet the
FK percolation transition coincides with the ferromagnet-paramagnet, and 
here we
investigate whether this is true for any other values of $p$.
Results of some test simulations performed
previously \cite{hasenbusch2007} suggest this is not the case at least for
some values of $p$.
Also, close to the ferromagnetic-spin glass boundary,
where frustration is lower than for the standard spin glass model which
has $p=1/2$, we investigate whether the Wolff algorithm performs
better than for the standard spin glass model. If
this were the case, one might be able to
study low-temperature spin-glass behaviour
for larger samples, by working in this range of $p$.

We present an extensive study of the FK percolation
transition in the full range of interest $0\le p\le 1/2$, 
which indicates that this
transition happens above the phase transition line
\textit{for all} $p>0$.
Only for the pure ferromagnet, $p=0$, does it coincides with the
FM-PM transition. In addition, the critical exponents seem to be those of the 
(uncorrelated) percolation problem
everywhere along the FK transition percolation transition
line, including both the ferromagnetic and
spin-glass regions, see Fig.~\ref{fig:phasediagram}. The only exception is for
$p$ \textit{precisely} equal to 0, 
the pure
ferromagnet, for which the critical exponents are those
of the Ising model.
Finally, our result indicate that in the spin glass region
close to $p_c$ the
Wolff algorithm does not perform notably better than for the standard ($p=1/2$)
spin-glass case.

Our paper is organised as follows. In Sec.~\ref{sec:methods}, we 
review the algorithms we have used. Next, in  Sec.~\ref{sec:results},
we present our
results, and finally in Sec.~\ref{sec:summary} we give a summary and discussion.

\section{\label{sec:method}Methods}
\label{sec:methods}

To study the FK percolation transition
and to investigate 
the efficiency of the Wolff
algorithm
we construct FK clusters at each step as follows: 
\begin{itemize}
\item
Bonds where $J_{i,j}\sigma_i \sigma_j>0$
are said to be \emph{satisfied}, and we \textit{activate} them
with probability $p_{\mathrm{act}}=1-e^{-2\beta
|J_{ij}|}$. Unsatisfied bonds are never activated.
\item  We determine all clusters of
spins connected by activated bonds, as in bond
percolation.
\end{itemize}
A cluster is said to be \emph{wrapping} or \emph{percolating} 
if it spans the lattice across between the
periodic boundaries and so is connected back to itself. 
For each step, we record whether a cluster is wrapping (this is typically the 
largest one), and we also monitor
the sizes of all clusters to investigate the distribution of
cluster sizes. Finally, we generate the next configuration according
the Wolff algorithm by selecting a spin at random and flipping the
the spins (i.e. with ``acceptance probability'' one) in the cluster 
which contains it.

Averages are done both over the spin configurations for a given
realization and a disorder average over a large number of different
realizations.
The quantities that we measure are:
\begin{itemize}
\item The average wrapping probability, $p_{\rm wrap}$.
\item The fraction of sites in the largest cluster, $P$.
\item The number $n_s$ of clusters of size $s$.
\item The average size $S$ of the clusters excluding the
largest one (this would be the percolating cluster in the percolating phase).
The average
is done with respect to all sites,
i.e.~$S = \sum_s s^2n_s/\sum_s sn_s$. 
\item The average size of the flipped clusters, $n_{\rm Wolff}$.
\end{itemize}

For high temperatures the activation probability $p_{\mathrm{act}}$
is small, leading to many small clusters which do not wrap. On the
other hand, for low temperatures, $p_{\mathrm{act}}$ will be large
leading to few clusters and typically one big wrapping cluster. Thus,
in between, there exists a percolation transition of the FK 
clusters at some temperature 
$T_{\rm FK }$,
such that, in the thermodynamic limit, $N\to\infty$,
one finds $p_{\rm wrap}\to 0$
for $T>T_{\rm FK}$ and  $p_{\rm wrap}\to 1$
for $T<T_{\rm FK}$.

We analyse our data using finite-size scaling (FSS), as is
standard in percolation transitions.\cite{stauffer1994introduction}
According to FSS, at a
second order percolation transition 
near the critical point, the wrapping probability
should  exhibit a scaling behaviour
\begin{equation}
p_{\rm wrap}(L,T) = f_{\rm wrap}((T-T_{\rm FK})L^{1/\nu})\,,
\label{eq:fss:wrapping}
\end{equation}
where $\nu$ is the critical exponent which describes the divergence
of the correlation length of the FK clusters. Thus, 
the parameters $T_{\rm FK}$ and $\nu$ can be determined by varying them 
until the data for different
system sizes collapse on to the same universal curve $f_{\rm wrap}(\tilde x)$.

Furthermore, in the percolating phase, the fraction of sites in the largest
(i.e. percolating) cluster in an infinite system goes to zero like
$P \sim (T_{\rm FK} -T)^{\beta}$ as $T$ approaches $T_{\rm FK}$ from
below.  For a finite system, this becomes, according to FSS,
\begin{equation}  
P(L,T) = L^{-\beta/\nu}f_{P}((T-T_{\rm FK})L^{1/\nu})\,,
\label{eq:clustersize}
\end{equation}
allowing us to obtain the critical exponent $\beta$. The 
average cluster size  behaves in a similar way, as described
by the finite-size scaling relation 
\begin{equation}
S  (L,T) = L^{\gamma/\nu}f_{S}((T-T_{\rm FK})L^{1/\nu})\, .
\end{equation}
Note that in computing $S$ we neglect the largest cluster, so $S$ has a
maximum near the 
percolation transition, because in the non-percolating
phase there are only many small clusters, while in the percolating phase
most sites belong to the percolating cluster which is neglected.
Thus, 
the scaling function $f_{\rm S}$  exhibits a peak at some value $x_{\rm peak}$,
corresponding to a temperature $T_{\rm peak}=T_{\rm FK}+x_{\rm peak}L^{-1/\nu}$,
which means that the height of $S^\star$ at the peak scales with a power-law
\begin{equation}
S^\star \sim L^{\gamma/\nu}\, ,
\label{Sstar}
\end{equation}
allowing us  to obtain the critical exponent $\gamma$. Finally, at the
critical point $T_{\rm FK}$,
the distribution $n_s$ of cluster sizes for an infinite system is expected
to follow a power-law 
\begin{equation}
n_s(T_{\rm FK})\sim s^{-\tau}\, ,
\end{equation}
defining another critical exponent $\tau$.

The critical exponents are not independent of each other. Instead,
they are connected through scaling relations, such that 
there are only two independent exponents.
The scaling relations for the 
standard percolation problem are often
expressed \cite{stauffer1994introduction} as functions of
exponents describing the shape of $n_s$, i.e.~for an infinite system
\begin{equation}
n_s = s^{-\tau} f_n\left( s^\sigma\, (T-T_{\rm FK}) \right) \, ,
\end{equation}
which defines another exponent $\sigma$.
In terms of $\tau$ and $\sigma$
the standard scaling relations are \cite{stauffer1994introduction}
\begin{equation}
\label{eq:beta}
\nu=\frac{\tau-1}{\sigma d}\,,\;
\gamma=\frac{3-\tau}{\sigma}\,,\;
\beta=\frac{\tau-2}{\sigma}\,.
\end{equation}

We don't measure $\sigma$, since this would require additional
numerical effort, but we
can remove $\sigma$ from the equations by solving the 
first equation with respect to $\sigma$ and inserting the solution
into the other two, resulting in:
\begin{equation}
	\gamma=\frac{3-\tau}{\tau-1}\nu d\,,\;
	\beta=\frac{\tau-2}{\tau-1}\nu d\,.
\label{eq:scaling:relations}
\end{equation}

We will verify that our computed
values for $\nu,\tau,\gamma$ and $\beta$ 
obey these relations.

\section{Results}
\label{sec:results}

We perform simulations for various values of $p\in [0,0.5]$. For 
each value of $p$ we treated different system sizes $L \in [10,100]$,, and for a
few values of $p$ we also did simulations for $L=200$, see below.
All results are disorder averages over typically 1000 realisations.
For each realisation  we perform Monte Carlo simulations using the
Wolff algorithm for 
72 temperatures 
equally spaced in $[3.615,4.68]$, i.e.~with spacing $\Delta T = 0.015$. 
For the selected cases of $p=0.1$, $0.3$ 
and $0.5$ (and also for $p=0$ as a comparison with other work and a check
on our code),
we studied 20 additional
temperatures spaced by $\Delta T=0.003$ very close to $T_{\rm FK}$,
in order to determine the critical properties precisely. 

To check for equilibration we
average over intervals $[t/2,t]$ for a logarithmically increasing set
of times $t$, and require that there is no systematic trend for the last
several values of $t$.
Typically, due to the high temperatures,
equilibration is achieved within a few steps. 
For small systems, $L\le 30$, we perform 2$\times 10^5$ Wolff
steps per realisation, while for the larger systems, which run
slower but still need only a few steps to equilibrate, we do
$5\times 10^3$ steps. 

\begin{figure}[ht]
\includegraphics[width=\columnwidth]{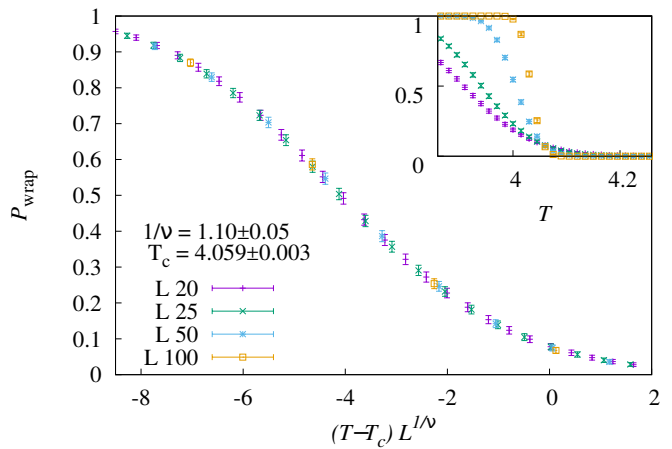}
\caption{\label{fig:kollaps_nu} Wrapping probability as function of
temperature $T$ for $p=0.1$, for various system sizes $L$.
The inset shows the raw data, while in the main plot a data collapse 
to determine $T_{\rm FK}$ and $\nu$ gives $1/\nu=1.10(5)$ and
$T_{\rm FK}=4.059(3)$.}
\end{figure}

To determine the position of the FK percolation transitions, 
we monitor the wrapping probability
of the FK clusters. An example is shown for $p=0.1$ in the inset
of Fig.~\ref{fig:kollaps_nu}. A clear decrease of the wrapping
probability beyond $T\approx 4$ is visible. We performed a data
collapse according to Eq.~(\ref{eq:fss:wrapping}), see main plot
of Fig.~\ref{fig:kollaps_nu}, to determine $T_{\rm FK}$
and the critical exponent $\nu$ of the percolation length, resulting in
$T_{\rm FK}= 4.059(3)$ and $\nu=0.89(8)$. 
The best fit parameters were determined from the method discussed in
the appendix of Ref.~[\onlinecite{gauss_2d}] and in
Ref.~[\onlinecite{melchert:09}].

In a similar way, we analysed the data for other values of $p$.
The resulting values of $T_{\rm FK}$ as a function of $p$ are shown in the
phase diagram in
Fig.~\ref{fig:phasediagram},
along with the values for the FM-PM and SG-PM
phase transitions obtained from the literature \cite{reger1986,hasenbusch2007},
and the critical concentration $p_c$ for the zero-temperature
FM-SG transition.\cite{art_threshold1999} Interestingly,
the FK percolation transition seems to coincide with magnetic-ordering
transition only for
the pure ferromagnetic system ($p=0$).
For all other values of $p$,
$T_c<T_{\rm FK}$ even close to the pure
ferromagnet. Hence, even if the ground state is ferromagnetic,
i.e.~for $0<p<p_{\rm c}$, the FM-PM phase transition cannot be understood as a
percolation transition of the FK clusters.

\begin{figure}[htb]
\includegraphics[width=\columnwidth]{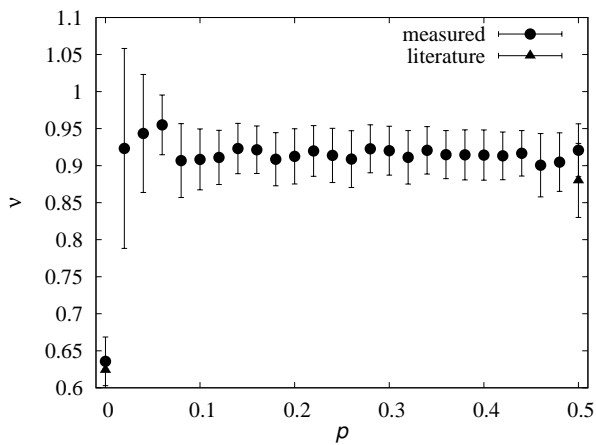}
\caption{\label{fig:nu} The critical exponent $\nu$ as a function of $p$. 
The value of $\nu$ for $p=0$ is from Ref.~[\onlinecite{baillie1992}] and the 
value for $p=0.5$ indicated by a triangle is
from Ref.~[\onlinecite{deArcangelis1991}].
}
\end{figure}

The resulting values of $\nu$ as a function of $p$ are shown in
Fig.~\ref{fig:nu}. For $p=0$, we recover the literature value for
the pure Ising ferromagnet,\cite{baillie1992} 
but with larger error bars (which is natural,
because our main numerical effort goes into the necessary disorder
average and considering several values of $p$).
For all other values of $p$, including
both ferromagnetic and spin glass regions,
we find that $\nu$ is compatible with
the previously found\cite{deArcangelis1991} value of $\nu=0.88(5)$. This 
is also compatible with the value\cite{wang2013}
for the standard percolation problem,
in which there are no
correlations between the occupancies of the bonds. 
By contrast, in FT clusters there
\textit{are}
correlations for all $p$ but interestingly they do not seem to affect the
critical behavior, except for $p=0$ where the bond occupancies are
\textit{rigorously constrained} to follow  Ising correlations.


To investigate universality more carefully we have evaluated the other
critical exponents with additional data near $T_{\rm FK}$ for the
values $p=0$ (for a consistency check), $p=0.1$ (a ferromagnetic case),
$p=0.3$ and $p=0.5$ (SG cases; for the latter value the critical behavior
is already partially known\cite{deArcangelis1991}).

\begin{figure}
	\includegraphics[width=\columnwidth]{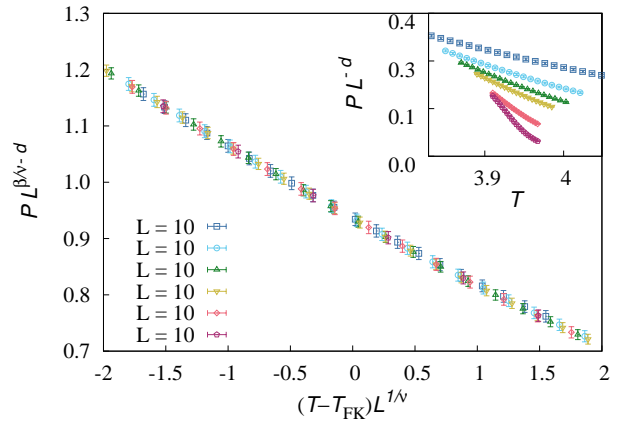}
	\caption{\label{fig:max_cluster_size} (color online) 
The fraction of sites in the infinite cluster, $P$,
as a function of the temperature $T$ in the vicinity of $T_{\rm FK}$, 
for $p=0.3$ and various system sizes $L$. 
The inset shows the raw data, while the main plot shows the
data rescaled according to Eq.~(\ref{eq:clustersize}), with
best fitting values $\beta=0.48(4)$, $\nu=0.87(8)$, and $T_{\rm FK}=3.939(3)$.}
\end{figure}

For the fraction of sites in the infinite cluster
(the order parameter), we show data for $p=0.3$
in
Fig.~\ref{fig:max_cluster_size}.
From a finite-size scaling collapse of the data we obtain the
best fitting parameters
$\beta=0.48(4)$, $\nu=0.87(8)$ and $T_{\rm FK}=3.939(3)$. The result
for the other intensively studied cases are shown in Table~\ref{tab:exponents}. 
Note that for $T_{\rm FK}$ and $\nu$, we usually have several independent 
estimates
available and the stated values and their error bars 
are chosen such that they are compatible with all results.

\begin{table*}[htb]
\caption{\label{tab:exponents}
Best estimates 
for the critical temperatures $T_{\rm FK}$ and critical exponents $\nu$,
$\tau$, $\gamma$ and $\beta$. The number in brackets denote the error
bars in the last digit. Also shown are the
values for $\gamma$ and $\beta$ obtained by inserting the values for $\nu$ and
$\gamma$ into the scaling relations in
Eq.~(\ref{eq:scaling:relations}). They agree with the values for $\gamma$ and
$\beta$ obtained directly within the error bars.
}
\begin{ruledtabular}
\begin{tabular}{llllllll} 
$p$ & $T_{\rm FK}$ & $\nu$ & $\tau$ & $\gamma$ & $\beta$
& $\gamma=\frac{3-\tau}{\tau-1}\nu d$ & $\beta=\frac{\tau-2}{\tau-1}\nu d$\\
\hline 
0.0 & 4.5116(5) & 0.65(4)& 2.27(3) & 1.18(6) & 0.31(4) & 1.1(2) & 0.4(1) \\ 
0.1 & 4.059(3) & 0.89(8)& 2.196(8)& 1.82(8) & 0.48(4) & 1.79(12)& 0.44(4)\\ 
0.3 & 3.941(3) & 0.89(8)& 2.23(5) & 1.84(6) & 0.41(4) & 1.7(2) & 0.49(16)\\ 
0.5 & 3.934(3) & 0.88(9)& 2.26(1) & 1.8(1) & 0.41(5)  & 1.6(2) &  0.54(9)\\ 
\end{tabular}
\end{ruledtabular}
\end{table*}

\begin{figure}[htb]
\includegraphics[width=\columnwidth]{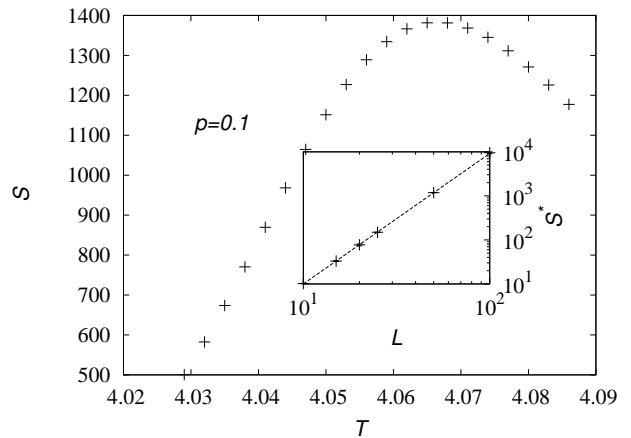}
\caption{\label{fig:mean_cluster_size} 
Mean cluster size $S$ at $L=50$ 
and $p=0.1$ as a function of temperature $T$, near the FK 
percolation transition $T_{\rm FK}$. The data exhibits a peak with peak
height $S^\star$. The inset shows the peak height as function of $L$.}
\end{figure}

In addition to the order parameter, we have also ana\-lysed the data for
the average cluster size $S$. As example, we show the result for
$p=0.1$ and size $L=50$ as a function of temperature $T$
in Fig.~\ref{fig:mean_cluster_size}. The data exhibits a peak
at some point ($T^\star, S^\star$). One can 
read off the critical exponent $\gamma$ from the 
$L^{\gamma/\nu}$ scaling, see Eq.~(\ref{Sstar}),
of the peak height as a function of $L$. The data is shown in
the inset of Fig.~\ref{fig:mean_cluster_size}.
For different values of $p$, the resulting values of $\gamma$ are also shown in 
Table~\ref{tab:exponents}. Again, we observe that for $p>0$
the results seem to agree with each other.

\begin{figure}[htb]
\includegraphics[width=\columnwidth]{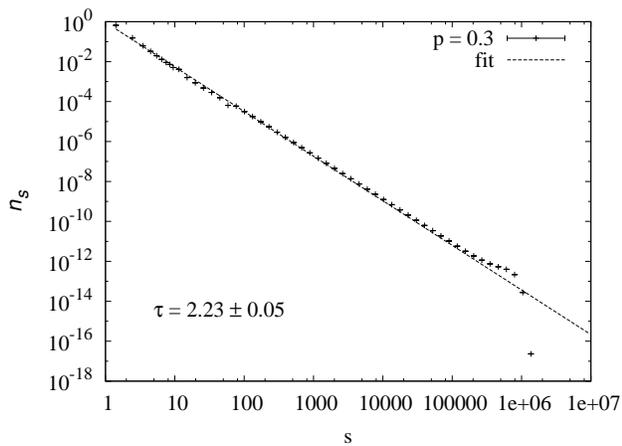}
\caption{\label{fig:tau}Cluster size distribution at $p=0.3$ with a system size of $N=200^3$.}
\end{figure}

To obtain the critical exponent $\tau$, we analyse the
distribution of cluster sizes, excluding the largest cluster,
at the critical point for a rather large system
size, $L=200$. As an example, we present our results for $p=0.3$
in Fig.~\ref{fig:tau}. The data exhibits a high
quality which allows us to observe a power law over about 10 decades
in probability. A fit resulted in a value $\tau=2.23(5)$. This value,
and the results for the three other selected cases, are also shown in 
Table~\ref{tab:exponents}. The values we have found for all values of $p>0$
are compatible
with the values for standard percolation in three dimensions.

The exponents should obey the scaling relations
in Eqs.~(\ref{eq:scaling:relations}). The values obtained when inserting
the measured values for $\nu$ and $\tau$ from Table~\ref{tab:exponents} to
estimate $\gamma$ and $\beta$ from the scaling relations are shown in the last two
columns in
Table~\ref{tab:exponents}. All these values 
are compatible with the directly measured values within error bars.
Note that the error bars from the scaling relations
are larger than the error bars of the directly measured
exponents due to error propagation.

\begin{figure}[htb]
\includegraphics[width=\columnwidth]{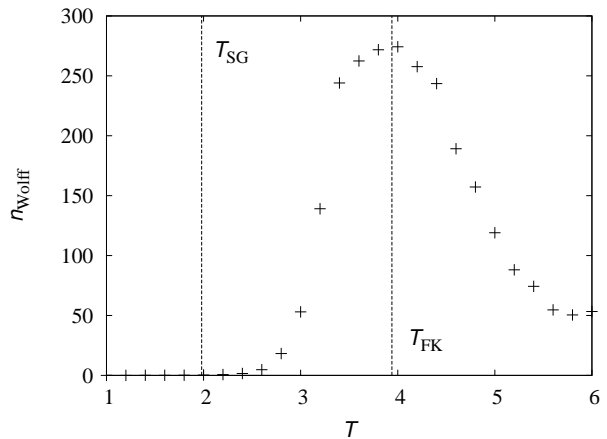}
\caption{\label{fig:gzb} Average size of the clusters flipped (or not
flipped if this is smaller) by 
the Wolff algorithm as a function of temperature $T$ for $L=10$ and 
$p=0.25$.}
\end{figure}

Finally we consider the question of whether the Wolff algorithm
might be more efficient in the spin-glass phase near the
FM-SG transition, i.e.~for $p$ just slightly greater than $p_c$, rather than
for $p=1/2$.
In Fig.~\ref{fig:gzb} we show the average effective 
size of the
flipped cluster (which is not always the largest one), as a function
of the temperature $T$ for $p=0.25\, (>p_{\rm c})$. By ``effective'' we mean that
if the cluster of flipped spins is larger than half of the system
size, then the spins which are not flipped are counted.
We see that the clusters which are flipped near the SG phase transition are very small.
One could already expect this from the
phase diagram in Fig.~\ref{fig:phasediagram}, which shows that for $p=0.25$
the FK percolation transition is considerably above the critical temperature
$T_{\rm SG}$. Thus, applying the Wolff algorithm in the spin glass phase but 
near the spin glass-ferromagnet phase boundary, 
does not lead to any benefit relative to studying the standard spin glass
model which has $p=1/2$.

\section{Summary} 
\label{sec:summary}

We have studied the percolation transitions of Fortuin-Kasteleyn
clusters for the three-dimensional random-bond Ising model.
Near the cluster percolation transition
the Wolff algorithm can be used to
efficiently sample equilibrium configurations.
However, except for the pure Ising case ($p=0$), the
temperature of the percolation transition is higher that of the
ferromagnet-paramagnet and spin glass-paramagnet transitions, 
and for most values of $p$ it is \textit{much} higher, see
Fig.~\ref{fig:phasediagram}. 
This 
renders the Wolff algorithm inefficient for the magnetic
transitions except for $p=0$.  
Indications of this behaviour were already found in
some test simulations of a previous study,\cite{hasenbusch2007}
where, for the FM-PM 
phase boundary at one value of $p>0$, cluster algorithms were tried but
turned out to be inefficient.

We have determined the critical exponents at the FK cluster percolation transition.
For $p=0$, the pure Ising case, we obtain the known values, which are those of
the Ising model since the FK clusters are controlled by Ising correlations in
this limit.
For all other values, $0<p\le 1/2$, our results are
compatible with the universal behaviour of standard percolation, 
irrespective of whether the ground
state exhibits ferromagnetic or spin-glass order. Since standard percolation
has no correlations between the occupancy of the bonds, whereas bonds in the
FK clusters \textit{are} correlated, this implies that the correlations are irrelevant
for universal properties, and so presumably are of short range for $p > 0$. 


For future studies, it would be interesting to investigate other
types of cluster algorithms \cite{niedermayer1988,joerg2005,zhu2015}
for the three-dimensional random-bond Ising model.
So far, from the literature studies
known to us, none of them turned out to be
efficient enough to study the pure spin-glass case ($p=1/2$)
for large enough systems, but
it could be that some will work well close to the FM-SG phase boundary or 
perhaps
at least for ferromagnetic ordering of the random ($p>0$) case. 

\section{Acknowledgements}

APY thanks the Alexander von Humboldt Foundation for financial support
through a Research Award.
The simulations were performed at the HPC facilities GWDG G\"ottingen, HERO and
CARL. HERO and CARL are both located at the University of Oldenburg (Germany)
and funded by the DFG through its Major Research Instrumentation Programme
(INST 184/108-1 FUGG and INST 184/157-1 FUGG) and the Ministry of Science and
Culture (MWK) of the Lower Saxony State.

\bibliography{references}
\end{document}